\def \be {\begin{equation}}
\def \ee {\end{equation}}
\def \bea {\begin{eqnarray}}
\def \eea {\end{eqnarray}}

\def \la {\langle}
\def \ra {\rangle}
\def \R {{\bf R}}

\def \dels {\partial\kern-.5em / \kern.5em}
\def \As {{A\kern-.5em / \kern.5em}}
\def \Ds {D\kern-.7em / \kern.5em}

\def \a {\alpha}

\def \eps {\epsilon}

\def \lam {\lambda}
\def \Lam {\Lambda}
\def \s {\sigma}

\def \one {{\bf 1}}
\def \th {\theta}

\def \p {\partial}

\def \H {{\cal H}}

\documentclass[12pt]{article}

\begin{document}
\begin{titlepage}

\begin{center}
\hfill hep-th/0004072\\
\vskip .5in

\textbf{\large Fuzzy Spheres in AdS/CFT Correspondence \\
and Holography from Noncommutativity }

\vskip .5in
{\large Pei-Ming Ho$^1$, Miao Li$^{2,1}$}
\vskip 15pt

{\small \em $^1$ Department of Physics, National Taiwan
University, Taipei 106, Taiwan}

\vskip 15pt
{\small \em $^2$ Institute of Theoretical Physics, Academia Sinica,
Beijing 100080}

\vskip .2in
\sffamily{
pmho@phys.ntu.edu.tw\\
mli@phys.ntu.edu.tw}

\vspace{60pt}
\end{center}
\begin{abstract}
We show that the existent fuzzy $S^2$ and $S^4$ models are natural
candidates for the quantum geometry on the corresponding spheres in
AdS/CFT correspondence. These models fit nicely the data from
the dipole mechanism for the stringy exclusion principle.
In the $AdS_2\times S^2$ case, we show that a wrapped fractional
membrane can be used to count for the large ground state degeneracy.
We also propose a fuzzy $AdS_2$ model whose fundamental commutation
relation may underlie the UV/IR connection.

\end{abstract}
\end{titlepage}
\setcounter{footnote}{0}

\section{Introduction}

The nature of spacetime in string/M theory is becoming increasingly
discernible. The most universal, yet quite qualitative property
appears to be the spacetime uncertainty principle \cite{UNC}. 
This principle is in the same spirit as the UV/IR connection in 
the AdS/CFT correspondence \cite{UVIR}, the latter underlies most
of physics in by far the only concrete realization of the holographic
principle. Spacetime noncommutativity manifests itself in many
different situations in string/M theory. Its physical origin is
the fact that all physical objects in the theory are inherently
extended objects, such as strings, D-branes, and D0-brane partons
in matrix theory \cite{MT}. It is not surprising therefore that
when one increases one's time resolution, any probe available will
increase its physical size, and the space becomes more and more uncertain.
This fundamental property of string/M theory is counter to the
usual Wilsonian renormalization group intuition. It may be the ultimate
key to resolving some long-standing puzzles, such as the quantum
black hole physics, the problem of cosmological constant.  

There exists concrete mathematical realization of space noncommutativity. 
When D-branes are embedded
in a constant B field background, the world-volume theory becomes
noncommutative, and this subject has recently attracted a lot of
attention. Yet this noncommutativity is best viewed as effective only,
although its origin is also stringy. In our view, the recently
proposed mechanism to explain the ``stringy exclusion principle''
\cite{SEP} hints at an even more interesting possibility, that we
can explore the spacetime noncommutativity in the full string/M
theory. There have
been proposals that the remarkable phenomenon of ``stringy exclusion''
is due to noncommutativity
of the quantum sphere in question \cite{JR,HRT}. The mechanism of
\cite{MST} provides a physical means for us to directly study 
the nature of the sphere in AdS/CFT correspondence. It was already
pointed out in \cite{MIAO} that this mechanism agrees very well with
the spacetime uncertainty relations, and a better model for noncommutativity
on the sphere is the fuzzy sphere. It naturally respects the rotational
invariance.
We will find that the data coming from quantizing a dipole a la Myers
\cite{RM} mesh perfectly with fuzzy spheres whose construction is
available. One might say that the noncommutative Yang-Mills theory
is a geometric manifestation of the B field
on the D-brane worldvolume,
and the fuzzy sphere is
a geometric manifestation of a R-R anti-symmetric tensor field
on the spacetime.

We will in the second section discuss the $AdS_2\times S^2$ case.
The dipole formed of a wrapped membrane is studied, we find that
its location on $S^2$ is quantized. This is the origin of the fuzzy
$S^2$. It is quite novel that for angular momentum $M\le N$, $N$ the
cut-off, the membrane is static on $S^2$, so the angular momentum 
is completely
induced from the Chern-Simons coupling. We also argue for the existence
of a fractional membrane whose tension is only $1/N$ of its original
tension. This fact together with the cut-off on the angular momentum
enables us to count the ground state entropy of $AdS_2\times S^2$.
It is of the order $N^2$. One may view our explanation as a bulk
microscopic one for the entropy. We also propose a fuzzy $AdS_2$ model.
Interestingly, the noncommutativity between time and the radial coordinate
may be regarded as a fundamental explanation of the UV/IR relation.
In sect.3, we show that the fuzzy $S^4$ constructed in
\cite{CLT} fits nicely the data obtained in the dipole mechanism.
Finally in sect.4, we motivate these fuzzy sphere models by considering
matrix theory on a sphere.
 
The construction of fuzzy spheres of 2 and 4 dimensions in this paper
can be generalized to any even dimension.
However, it does not work for odd dimensions,
so our proposal is not contradictory to the proposal of
q-deforming $AdS_d\times S^d$ spaces for $d=3,5$ \cite{JR,HRT}.

Surely there are many questions that remain, such as how to use fuzzy
spheres to do physics directly. We hope to return to these problems
later.

\section{$AdS_2\times S^2$ and Fuzzy Sphere}

\subsection{The Dipole Mechanism} \label{dipole}

Among all examples in AdS/CFT correspondence, the boundary
theory of $AdS_2\times S^2$ is most poorly understood \cite{ANDY}. 
However, since the ``stringy exclusion principle" works in all other cases,
we believe it still holds here in this case. What we shall
say in this section is somewhat speculative, but the fact that
we can accurately account for the ground state degeneracy  lends
a strong support to this picture.

$AdS_2\times S^2$ can be obtained by taking the near horizon
limit of the 4 dimensional extremal Reissner-Nordtstr\"om
solution. The metric reads
\bea \label{adsm}
&ds^2=l_p^2\left(-{r^2\over N^2}dt^2+{N^2\over r^2}dr^2
+N^2d\Omega_2^2\right),\\
&F=-N d\Omega_2,
\eea
where $l_p$ is the 4 dimensional Planck length, $N$ is the magnetic
charge, and integrally quantized. This metric can also be obtained
by taking the near horizon limit of the 4 dimensional charged
black hole in string theory \cite{MS}. Unavoidably, there must be
four different charges $Q_i$, each associated to a kind of branes.
For instance, by wrapping two sets of membranes and two sets of
M5-branes in $T^7$, one obtains a 4D charged, extremal black hole
\cite{KT}. The brane configuration is as follows. Denote the
coordinates of $T^7$ by $x_i$, $i=1, \dots, 7$. A set of membranes
are wrapped on $(x_1, x_2)$, another set are wrapped on $(x_3, x_4)$.
A set of M5-branes are wrapped on $(x_1, x_3, x_5, x_6, x_7)$,
the second set are wrapped on $(x_2, x_4, x_5, x_6, x_7)$.
By setting all charges to be equal $Q_i=N$, the above metric results,
and the magnetic field is just the linear combination of all
anti-symmetric tensor fields involved. Note that here for
simplicity,  we consider
the most symmetric case in which all the charges  appearing
in the harmonics $1+Q_il_p/r$ are just $N$ which in turn is equal
to the number of corresponding branes used to generate this 
potential. As a consequence, the tension of the branes compensates 
the volume of the complementary torus.
This means that the size of each circle of $T^7$ is at the scale
of the M theory Planck length.

The size of the torus $T^7$ is unchanged by going to the near
horizon limit. 
This fact makes the implementation of the dipole mechanism
of \cite{MST} a subtle problem. To understand this point, imagine
that a graviton moving on $S^2$ is the manifestation of
a membrane wrapped on $T^2$ of $T^2\times T^5$. This membrane
is charged with respect to the field generated by M5-branes
wrapped on the $T^5$ factor. Now the corresponding $F$ is the
reduction of $F^{(4)}$ with two indices along $T^2$. The coupling 
of the membrane to this field is therefore $\int C^{(3)}$.
Upon integrating over $T^2$, we obtain
\be
N\int \cos\theta \dot{\phi}dt,
\ee
where we used $d\Omega_2 =\sin\theta d\theta\wedge d\phi$.
Assuming the membrane move only along the $\phi$ direction
so that the kinetic term is
\be
-{1\over l_p}\int\sqrt{1-R^2\sin^2\theta\dot{\phi}^2}dt.
\ee
This term is too large and will make the total energy of 
the candidate graviton the Planck scale, as we shall see in a moment.

Fortunately, this problem is resolved by a mechanism to obtain
a fractional membrane.  The membrane transverse to a set of M5-branes
necessarily lies on $T^2$ parallel to another set of M5-brane.
If this membrane is tightly bound to these M5-branes, its tension is
down by a factor $1/N$, and its charge is unaltered. To see that its
tension indeed becomes much smaller, imagine that one of the circles
in $T^2$ is the M circle, then the membrane is interpreted as a
fundamental string, and the M5-branes to which it is bound are
interpreted as D4-branes. It is well-known that a fundamental string
melt into D-branes has a tension
$(g_s/N)$ times its original tension \cite{GKP}. 
Since the M circle in question is at the Planck scale, so $g_s=1$.
Thus the tension of the membrane is down by a factor $1/N$.
Now, the total action of the membrane is given by
\be
S=-{1\over R}\int \sqrt{1-R^2\sin^2\theta\dot{\phi}^2}dt
+N\int\cos\theta\dot{\phi}dt,
\ee
where we used $R=Nl_p$. 

The angular momentum conjugate to
$\phi$ is
\be
M=R(1-R^2\sin^2\theta\dot{\phi}^2)^{-1/2}\sin^2\theta\dot{\phi}
+N\cos\theta.
\ee
Upon quantization, $M$ is an integer. 
Solving $\dot{\phi}$ in terms of $M$ and substituting it into the
energy, we find
\be
E(\theta)={1\over R\sin\theta}\left(\sin^2\theta
+(M-N\cos\theta )^2\right)^{1/2}.
\ee
For a stable orbit, $dE/d\theta =0$. This condition leads
to
\be
(M-N\cos\theta)\left(N\sin^2\theta -\cos\theta (M-N\cos\theta )\right)
=0.
\ee
When $|M|\le N$, the solution is
\be \label{qcon}
\cos\theta ={M\over N}.
\ee
And in this case the energy $E=1/R$ which is independent of
$M$. This we take as a surprising result. Its origin is the fact that
the membrane is completely at rest at the angle $\cos\theta =M/N$, its
angular momentum is induced from the C-S coupling.

When $|M|>N$, the other solution is
\be \label{larg}
\cos\theta ={N\over M},
\ee
and the energy is
\be
E={1\over R}\sqrt{1+M^2-N^2}.
\ee
In this case $E$ does depend on the angular momentum $M$.
Unlike in its higher dimensional analogue \cite{MST}, the
dipole mechanism alone does not forbid higher angular momentum.
(It can be checked that at values in (\ref{qcon}) and (\ref{larg}),
the energy is always a local minimum.)

It is interesting to note that for the standard membrane whose
tension is not fractional (this membrane is not bound to
a set of M5-brane), $|M|\le N$ modes saturate
the M theory spacetime uncertainty relation
$\Delta t\Delta x^2 >l_p^3$.
A calculation similar to the above leads to energy $l_p^{-1}$. 
The uncertainty in time is then
$\Delta t\sim l_p$. The uncertainty in space is $\Delta x\sim l_p$,
since it is just the size of a circle.
Of course for a fractionated membrane, the energy is smaller, thus
the uncertainty in time is larger, the spacetime uncertainty relation
is satisfied. It is also interesting to note that for 
the standard membrane with $|M|>N$, the uncertainty relation
is violated, and on this basis we rule out larger angular 
momenta.

\subsection{Fuzzy $S^2$}

The above discussion naturally leads us to a fuzzy sphere
model for the factor $S^2$ in $AdS_2\times S^2$. The definition
of fuzzy $S^2$ \cite{MADORE} is specified by a representation
of the $su(2)$ Lie algebra
\be \label{S2}
[X^i,X^j]=i\epsilon^{ijk}X^k.
\ee
This algebra respects $SO(3)$ invariance.
If the representation is irreducible and is $2N+1$ dimensional,
then the Casimir is 
\be
\sum_i (X^i)^2= N(N+1).
\ee
The eigenvalues of any of $X^i$ are $-N, \dots, N$. 

It is rather clear now that if we identify $(X^il_p)$ with
the Cartesian coordinates of $\R^3$ in which $S^2$ is embedded,
we get a nice match between
the fuzzy sphere and the physics we have learned. The radius
can be defined either by the largest of the eigenvalue of
$X^3$, say, or by the Casimir. The difference between these two
definitions becomes vanishingly small in the large N limit.
Now the eigenvalue of $X^1 l_p/R$ are $M/N$, with $|M|\le N$.
Using the polar coordinates, this is just the quantization
on $\cos\theta$ in (\ref{qcon}). The algebra generated
by $X^i$ is $(2N+1)^2$ dimensional.
This happens to be
equal to the number
of all modes of angular momenta satisfying 
$M\le 2N$:
\be \label{mult}
\sum_{M=0}^{2N} (2M+1)=(2N+1)^2.
\ee

This result is reminiscent of the stringy exclusion principle
in $AdS_3\times S^3$ \cite{SEP}.
In the CFT dual to $AdS_3\times S^3$ there are a number of
interesting properties \cite{SEP,JR}.
First, the chiral primary operators of charges higher than $(N, N)$
for the quantum numbers $(2J_L, 2J_R)$ with respect to
the $SU(2)_L\times SU(2)_R$ symmetry group
can be written in terms of products of chiral primary operators
of lower charges.
Since products of chiral primary operators in the dual CFT
are interpreted to correspond to multi-particle states
in $AdS$, this means that there is a cutoff on the angular momenta
by $N$ for single-particle states.
The stringy exclusion principle is that multi-particle states 
also have a cutoff of angular momenta at $2N$.
It is natural to adopt the same interpretation here for $S^2$.
The algebra generated by $X$ contains multi-particle states
and has the cutoff of angular momentum at $2N$ as shown in (\ref{mult}).
The single-particle states is cutoff at angular momentum $N$
as we have shown earlier in sect.\ref{dipole},
which is also the angular momentum of
the $2N+1$ dimensional representation of $X$.

To compare with the construction of fuzzy $S^4$ we will discuss
later, we point out that there is another representation of
$X^i$ based on Pauli matrices $\sigma_i$. Let $X^i$ be
the following matrices acting on the tensor space $V^{\otimes n}$
where $V$ 
is a two dimensional vector space:
\be \label{tenp}
X^i=\left (\sigma_i\otimes \dots \otimes 1_2 +\dots
+1_2\otimes\dots \otimes\sigma_i\right)_{\mbox{sym}},
\ee
where the subscript `sym' indicates the totally symmetrized tensor 
product of $\otimes V$. If the rank $n$ of this tensor product is $2N$, 
obviously the totally symmetrized space is $2N+1$ dimensional,
and the matrices $X^i$ are $(2N+1)\times (2N+1)$ matrices.
These matrices satisfy the $su(2)$ Lie algebra. We shall see
that the fuzzy $S^4$ we need is a simple generalization of
the construction (\ref{tenp}). Moreover, the rank of the tensor product
is also $2N$.

Finally, we note that on the fuzzy $S^2$, $\cos\theta$ is conjugate
to $\phi$ with the ``Planck constant'' given by $1/N$, namely
$\cos\theta =-(i/N)\partial_\phi$,
because the angular momentum $M$ is conjugate to $\phi$.
Using this and the identification
\be
X_1=N\sin\th\cos\phi,\quad X_2=N\sin\th\sin\phi, \quad X_3=N\cos\th,
\ee
one can in fact derive the algebra of fuzzy sphere (\ref{S2}).

\subsection{The Entropy of $AdS_2\times S^2$}

An interesting property making $AdS_2\times S^2$ quite
different from other AdS spaces is its huge ground state degeneracy.
The entropy is simply $\pi N^2$, a quarter of the area of
$S^2$ measured in the Planck unit.
Its origin is the Bekenstein-Hawking entropy of the 4D charged
black hole. Of course this entropy was explained microscopically
using the brane construction \cite{MS}, although as any such
explanation, the argument goes through when the brane theory is
weakly coupled, thus the horizon is not macroscopic.

As the dual theory of $AdS_2\times S^2$ is not known yet, the best
one can do is to give an bulk explanation of the entropy of the
ground states. That one can do this is one advantage of working
in low dimensions. We have argued that the angular momentum
has a cut-off $N$, so the gravity theory in the full four dimensional
spacetime reduces to one in two dimensions only. One does expect
that the resulting 2D theory is renormalizable. In our argument
for the entropy we will ignore interactions, so we are dealing with
a free theory on $AdS_2$. Consider a finite temperature situation.
The relevant Euclidean metric we will use is
\be
ds^2={R^2\over z^2}(dt^2+dz^2),
\ee
namely it is the metric on the upper half plane. The conformal
factor is irrelevant for a massless field. As we have seen,
all the 2D fields in question are massive, and their mass is
independent of angular momentum, and is just
$m=1/R$. Thus for a scalar field, the Euclidean action is
\be 
S_E={1\over 2}\int dtdz\left( (\p_t\phi)^2+(\p_z\phi)^2
+{1\over z^2}\phi^2\right).
\ee
The partition function is given by
\be
Z=\int [d\phi] e^{-S_E}.
\ee
Due to a nice property of $AdS_2$, the partition function is
independent of the temperature. This is because we can rescale
the period $\beta$ of the Euclidean time away by performing
$t\rightarrow \beta t$, $z\rightarrow \beta z$, and the action
remains intact. As a consequence, the partition function
is a pure number. As a bonus,
the average energy
\be
\la E\ra =-\p_\beta \ln Z=0,
\ee
thus we always stay in the ground state. Thus the contribution
to the entropy from a scalar is simply $S=\ln Z$. If there
are $N^2$ such scalars, the total entropy is
\be
S=N^2\ln Z.
\ee
Of course based on supersymmetry we also expect the contribution
from fermions. The number of fermions is also of the order $N^2$.
We conclude that the ground state entropy indeed is of the order
$N^2$. 
Note that the number of fields in $AdS$ induced from the fuzzy sphere
by Kaluza-Klein reduction is precisely proportional to $N^2$ at large $N$.
So we have accounted for the entropy in $AdS_2\times S^2$
up to an overall constant.

It remains to compute $Z$ exactly. The following simple argument
shows that it is a finite number. For a massive particle, 
$AdS_2$ acts as a finite box. Both the spatial size and the 
temporal size are the same, due to the scaling invariance.
The free energy $F$ is just the Casimir energy and scales inversely
with the size of the box, and consequently $\beta F$ is independent
of this size and is a finite number. Finally, there can be contribution
from other massive modes to the entropy, for instance contribution
from wrapped ordinary membranes. These membranes have energy of order
$l_p^{-1}$, much heavier, and we expect that their contribution is suppressed
by factor $1/N^2$.

\subsection{Fuzzy $AdS_2$}

The UV/IR connection has its simplest manifestation in $AdS_2$.
We already showed that the spatial coordinate $z$ scales with
time. Alternatively $r=1/z$ scales inversely with time, thus
acts as the energy scale.

We propose a simple explanation of this connection
based on fuzzy $AdS_2$.
For a massless field living on $AdS_2\times S^2$,
its decomposition into harmonic functions on $S^2$
must match its decomposition on $AdS_2$.
This requirement was used in \cite{JR} to argue that
if the $AdS_3$ part of $AdS_3\times S^3$ is q-deformed,
then the $S^3$ part should also be q-deformed.
For the same reason we expect that $AdS_2$ should be fuzzy
because $S^2$ is fuzzy.
Since the fuzzy $S^2$ is defined by saying that
the Cartesian coordinates satisfy the Lie algebra of $SU(2)$,
the fuzzy $AdS_2$ should be defined by saying that
the Cartesian coordinates satisfy the Lie algebra of $SU(1,1)$.

Let $X_{-1}, X_0, X_1$ be the Cartesian coordinates of $AdS_2$.
Then
\bea
&[X_{-1}, X_0]=-il_p X_1, \\
&[X_0, X_1]=il_p X_{-1}, \\
&[X_1, X_{-1}]=il_p X_0,
\eea
which is obtained from $S^2$ by a ``Wick rotation''
of the time directions.
The radial coordinate $r$ and the boundary time coordinate $t$
are defined in terms of the $X$'s as
\be
r=X_{-1}+X_1, \quad t=\frac{R}{2}(r^{-1}X_0+X_0 r^{-1}),
\ee
where we symmetrized the products of $r^{-1}$ and $X_0$
so that $t$ is a Hermitian operator. The metric in terms of these
coordinates assumes the form (\ref{adsm}).
It follows that the commutation relation for $r$ and $t$ is
\be \label{rt}
[r,t]=-iRl_p.
\ee

This relation (\ref{rt}) is suggestive of the identification of
$r/(Rl_p)$ with the conjugate variable of $t$,
which is just the Hamiltonian of the boundary theory.
This noncommutativity contains the UV/IR relation.
Furthermore, it also suggests the uncertainty relation
\be
\Delta r\Delta t\geq Rl_p,
\ee
which implies that if we demand to have a classical description of $t$,
then we will lose all physical distinction of the value of $r$.
This is just what holography is--
one can describe the theory in $AdS$ by a field theory
on the boundary space, which is viewed as a classical space.
It would be interesting to see if we can extend this nice
incorporation of holography in noncommutativity
to other examples of $AdS/CFT$ dualities.

The identification $E=r/(Rl_p)$ would seem a drastic reduction in
energy scale. This is not so. Indeed if one follows the analysis
in the second reference of \cite{UVIR}, one would find that for a 
massless graviton or a fractional membrane, the energy scale associated 
to $r$ is $E=r/R^2$,
even smaller than $E=r/(Rl_p)$. For a massive graviton whose mass
is the Planck scale, one would obtain our relation. This massive
graviton can be obtained by wrapping a membrane on $T^2$ of $T^7$.
However, as we have suggested, the modes responsible for the
ground state entropy appear to come from fractional membranes.

\section{Fuzzy $S^4$ in $AdS_7\times S^4$}

\subsection{Fuzzy $S^4$}

The first unambiguous implementation of the dipole mechanism
is in $AdS_7\times S_4$. The graviton moving on $S^4$ is polarized
under the $F^{(4)}$ field strength to become a membrane, by a
mechanism similar to one proposed by Myers \cite{RM}. The size
of the membrane is quantized \cite{MST, MIAO}, and it is natural
to conjecture that this quantization leads to a fuzzy $S^4$.

A fuzzy $S^4$ was proposed in \cite{CLT} to describe
a longitudinal M5 brane in the matrix theory.
Define the Cartesian coordinates of the fuzzy $S^4$
by totally symmetrized tensor products of
$n$ copies of gamma matrices as
\be \label{X}
X_i=\lam(\Gamma_i\otimes\one\otimes\cdots\otimes\one+
\one\otimes\Gamma_i\otimes\cdots\otimes\one+\cdots)_{\mbox{sym}}.
\ee
Thus $X_i$ are $4^n\times 4^n$ matrices.
Explicitly, we have
${(X_i)_{(a_1 a_2\cdots a_n)}}^{(b_1 b_2\cdots b_n)}$,
where $a_i, b_i=1,2,3,4$
are indices of a Dirac spinor in 5 dimensional Euclidean space.
The subscript `sym' in (\ref{X}) means that
all $a_i$'s and $b_j$'s are
separately totally symmetrized among themselves.
As an operator, $X_i$ can act on an element of $V^{\otimes n}$,
where $V$ stands for the space of 5D Dirac spinors.
Instead of symmetrizing the indices of $X_i$,
it is equivalent to restrict it to act on a smaller space $\H_n$,
which consists of only elements of $V^{\otimes n}$
which have totally symmetrized indices.
A basis of $\H_n$ is $\{v^{(k_1,k_2,k_3,k_4)}\}$,
where $v^{(k_1,k_2,k_3,k_4)}_{a_1\cdots a_n}$
equals one if there are $k_A$ indices $a_i$ equal to $A$,
and is zero otherwise.
(Obviously $\sum_{A=1}^{4}k_A=n$.)
The dimension of $\H_n$ is $N_0=\frac{1}{3!}(n+3)(n+2)(n+1)$,
hence $X_i$ can also be realized as $N_0\times N_0$ matrices.
The algebra generated by $X_i$ is invariant under $SO(5)$.

Let $\sigma_i$ ($i=1,2,3$) denote Pauli matrices.
We take the convention that
\bea
\Gamma_i&=&\sigma_i\otimes\sigma_3, \quad i=1,2,3, \label{Gamma} \\
\Gamma_4&=&\sigma_0\otimes\sigma_1, \\
\Gamma_5&=&\sigma_0\otimes\sigma_2, \label{G3}
\eea
where $\s_0$ stands for the $2\times 2$ unit matrix.
Therefore the index $a$ of a Dirac spinor can be written
as $a=(\a,\a')$ ($\a,\a'=1,2$) corresponding to
the decomposition of $\Gamma_i$ in (\ref{Gamma}).
One can check that
\be
\sum_{i=1}^{5}\Gamma_i\otimes\Gamma_i\simeq\one\otimes\one
\ee
when acting on $\H_2$.
Using this and $\{\Gamma_i,\Gamma_j\}=2\delta_{ij}$,
we can calculate
\bea
\sum_{i=1}^{5} X_i^2 &=& \sum_{i=1}^{5}
(\Gamma_i^2\otimes\one\otimes\cdots\otimes\one
+\one\otimes\Gamma_i^2\otimes\cdots\otimes\one + \cdots) \\
&&+2\sum_{A=1}^{n-1}\sum_{B=A+1}^{n}
(\one\otimes\cdots\Gamma_i\otimes\cdots
\otimes\Gamma_i\otimes\cdots\otimes\one) \\
&\simeq& (5n+n(n-1))\one\otimes\cdots\otimes\one,
\eea
where in the second line $\Gamma_i$ appears only
in the $A$-th and the $B$-th places in the tensor product.
It follows that
\be
R^2=\lam^2 n(n+4).
\ee
For large $n$, $\lam\sim R/n$.
For the four-form field flux $2\pi N$ on $S^4$,
the radius is $R=l_p(\pi N)^{1/3}$.

The identity
\be
\frac{1}{4!}\eps_{i_1 i_2\cdots i_5}X_{i_2}\cdots X_{i_5}=\lam^3 X_{i_1}
\ee
is used in \cite{CLT} to argue that this configuration
carries locally the charge for a longitudinal 5-brane.

What will be of interest is the spectrum of the operator
$\sum_{i=1}^{3}X_i^2$.
It is easier to calculate $\sum_{i=4,5}X_i^2$.
We have the identity
\be
\sum_{i=1}^{3}\sigma_i\otimes\sigma_i=2P-\one,
\ee
where $P$ is the permutation operator: $P(A\otimes B)=B\otimes A$.
So
\be
\sum_{i=4,5}\Gamma_i\otimes\Gamma_i=
2P_2-1-(\s_0\otimes\s_3)\otimes(\s_0\otimes\s_3),
\ee
where $P_2$ is defined by
\be
P_2((A_1\otimes B_1)\otimes(A_2\otimes B_2))
=((A_1\otimes B_2)\otimes(A_2\otimes B_1)).
\ee
Consider the following three vectors
$v^{(2,0,0,0)}, v^{(1,1,0,0)}, (v^{(1,0,0,1)}-v^{(0,1,1,0)})$ in $\H_2$.
Recall that an index $a$ has two parts $a=(\a,\a')$
in accordance with (\ref{Gamma}-\ref{G3}).
The first two vectors have their
second parts of indices $\a'$ symmetrized,
while the last vector has them antisymmetrized.
The spectrum of
$\sum_{i=1,2}(\s_0\otimes\s_i)\otimes(\s_0\otimes\s_i)$
is $\{0, 2, -2\}$, for the three vectors above, respectively.

Now consider the vector $v_m\equiv v^{(m,(n-m),0,0)}$.
For this case the first parts of indices are all equal to $1$,
and the second parts of indices are totally symmetrized.
It is straightforward to find that
$(\sum_{i=4,5}X_i^2)v_m=\lam^2(2n+4m(n-m))v_m$.
It follows that
\be \label{r2}
r^2\equiv\sum_{i=1}^{3}X_i^2=\lam^2\left( (2m-n)^2+2n \right)
\ee
when acting on $v_m$.
In the limit of large $n$,
\be \label{sinpsi}
\sin\psi\equiv\frac{r}{R}\simeq\frac{2m-n}{n}.
\ee

Another relation we will need below is
\be \label{X4X5}
[X_4, X_5]=2i\lam\sqrt{r^2-2\lam^2 n},
\ee
which approaches $2iRr/n$ in the large $n$ limit.

We do not claim that (\ref{r2}), (\ref{sinpsi}) and (\ref{X4X5})
supply the complete spectra for these operators.
We have only considered a special class of eigenvectors $v_m$.
As an example we consider the vector which is a linear
combination of vectors of the form $v^{(k_1,(n/2-k_1),k_3,(n/2-k_3))}$,
with half of its indices having their first parts being $1$,
and the other half being $2$.
We require that this vector has its second parts of indices
antisymmetrized between any two pairs of indices
which have different values for their first parts of indices.
(For a pair of indices whose first parts are the same,
their second parts are necessarily symmetrized.)
One finds that $\sum_{i=4,5}X_i^2\simeq 0$ on this vector.
This means that the maximal value of $r^2$ is $R^2$,
which is not included in (\ref{r2}).

\subsection{Noncommutativity of Fuzzy Graviton}

Following \cite{MST, MIAO}, consider a graviton moving in $S^4$.
According to Myers, the graviton is a fuzzy sphere due to
the $C$ field background,
so we have a 2-sphere in directions $X_{1,2,3}$ moving with
a constant angular momentum in the hemisphere parametrized by $X_{4,5}$.
Let the radius of the 2-sphere be denoted by $r$, and
\bea
&X_4=\sqrt{R^2-r^2}\cos\phi, \\
&X_5=\sqrt{R^2-r^2}\sin\phi,
\eea
where $R$ is the radius of $S^4$.
The angular momentum conjugate to $\phi$ is found to be $L=Nr/R$
for the giant graviton \cite{MST,MIAO}.
Upon quantization, we have the canonical commutation relation
$[L,\phi]=-i$, which implies that
$[r,\phi]=-i\frac{R}{N}$.
As a result, in the lowest order approximation (Poisson limit) in $1/N$,
\be
[X_4, X_5]\sim i\frac{R}{N}r.
\ee
Comparing this with (\ref{X4X5}), we find that we need $n=2N$.

As a further check, we recall that the stable value of $\sin\psi=r/R$
is found to be quantized as
\be
\sin\psi=M/N
\ee
for integer $M$ ranging from $0$ to $N$.
Comparing this with (\ref{sinpsi}),
we find a match if $m=M$ and $n=2N$. This match works in the
leading order in the large $N$ limit. This is not a worry, since
we expect corrections to $\sin\psi$ calculated in \cite{MST, MIAO},
just as we expect corrections to the energy formula $M/R$. For a scalar
graviton viewed in $AdS_7$, the correct energy formula is
$\sqrt{M(M+3)}/R$.

The fact that these membrane states do not give a complete
spectrum of $\sin\psi$ should not bother us because
they are not all the possible physical states.

It is also interesting to note that the commutation relations
among $X_{1,2,3}$ from (\ref{X}) are not exactly
those for a fuzzy 2-sphere.
It appears to be composed of two fuzzy hemispheres
with opposite orientations.
This again is not a contradiction because we can not
exactly identify the worldvolume noncommutativity of the membrane
with the target space noncommutativity of spacetime.
However, there should be some relations between the two kinds
of noncommutativities. For instance, one would intuitively
expect that the target space uncertainty should
not be smaller than the worldvolume uncertainty,
since the spacetime is itself defined by the probes.
There can be a closer relation between them but it is still elusive.

In \cite{BV} the Matrix model for the six dimensional
$(2,0)$ superconformal field theory was analyzed,
and it was found that the matrix variables for
the transverse coordinates of M5-branes
indeed satisfy an algebra which is essentially
the algebra of fuzzy $S^4$ (\ref{X}).

\section{Noncommutativity from Matrix Theory}

In this section we construct the matrix model
compactified on a $d$-sphere,
and show that the fuzzy $S^2$ and $S^4$ are
configurations with minimal energy.

The matrix model action in flat spacetime has
the potential term $\mbox{Tr}(\frac{1}{4}[X_i, X_j]^2)$,
where $X_i$'s are $N\times N$ matrices.
This term is invariant under the Poincare group.
It is natural to guess that
the matrix model compactified on a sphere
has a term like
\be \label{potential}
\mbox{Tr}\left(\frac{1}{4}[X_i, X_j]^2
+\frac{1}{2}\Lam(X_i^2-R^2)\right)
\ee
in the action,
where the index $i$ goes from $1$ to $d$ for a $d$-dimensional sphere,
and $\Lam$ is the Lagrange multiplier
by which the radius condition
\be \label{radius}
\sum_{i=1}^{d}X_i^2=R^2
\ee
is imposed.

New terms should be included in the matrix model
action in the presence of background fields.
For the case of $S^2$ and $S^4$, a three-form field background
with only indices in spatial directions exists.
It is coupled to a matrix current \cite{TAYLOR}
which vanishes when $\dot{X}_i=0$.
Since we will consider only the static states of the matrix model,
such interaction terms can be omitted.
The 6-form field background is also relevant for the case of $S^2$,
and the same thing happens \cite{TAYLOR}.

In \cite{DOUG} a matrix model for the 2-sphere is proposed
where the complex coordinates $z,\bar{z}$ are promoted to matrices.
However it was found that there is a singularity on the moduli space.
This corresponds to the loss of the global $SU(2)$ symmetry
due to the fact that this symmetry is nonlinearly realized
in terms of the complex coordinates, and thus can not
be preserved when $z,\bar{z}$ are noncommutative.
By adopting the Cartesian coordinates, we are able to preserve
the global symmetry, but the whole supersymmetric action
remains to be worked out.
For our purpose, (\ref{potential}) is the only part of the action
that we need.

The equations of motion
derived from the action (\ref{potential}) is
\be \label{XXX}
[X_j, [X_j, X_i]]-\Lam X_i=0,
\ee
in addition to the kinetic term,
which vanishes if we set $\dot{X}_i=0$ to minimize the total energy.
Thus the moduli space of this model is given by
solutions of (\ref{radius}) and (\ref{XXX}).
The fuzzy 2-sphere and 4-sphere, and its generalization to other
spheres of even dimensions following \cite{CLT},
are solutions of both relations.

Incidentally, an expansion of $X_i$ around the solution
of fuzzy spheres can be viewed as covariant derivatives
on dual fuzzy sphere.
There should be a T-duality which relates D0-branes
on a classical sphere to D2-branes on a fuzzy sphere.

What we have shown in this section is that
the static configuration of partons in the matrix theory
happens to coincide with the quantum geometry
which we argued to exist in some $AdS\times S$ spaces.
It would be interesting to further explore whether
this is a pure coincidence or a general rule.
Such rules should tell us about how to
identify the rank $N$ in the matrix theory
and the flux $N$ of background fields.

\section*{Acknowledgment}

We would like to thank D. Minic, S. Ramgoolam and A. Strominger for 
correspondence, and Y.S.~Wu for discussions.
This work is supported in part by
the National Science Council, Taiwan, 
and the Center for Theoretical Physics at National Taiwan University.
The work of M.L. is also
supported by a ``Hundred People Project'' grant.

\eject

\end{document}